# Tunable Itinerant Ferromagnetism in the Two-Dimensional FePd$_2$Te$_2$ Hosting 1D Spin Chains

Authors: Alberto M. Ruiz†, Andrei Shumilin†, Sourav Dey†, Diego López-Alcalá† and José J. Baldoví†,*

†Instituto de Ciencia Molecular, Universitat de València, Catedrático José Beltrán 2, 46980 Paterna, Spain. E-mail: j.jaime.baldovi@uv.es

**ABSTRACT**:

One-dimensional (1D) magnetism offers unidirectional spin interactions that allow unique tunable properties and unconventional spin phenomena. However, it often suffers from poor stability, limiting practical applications. In this regard, integrating 1D magnetism into two-dimensional (2D) materials enables a promising route to stabilize these systems while preserving their anisotropic magnetic characteristics. Here, we focus on the 2D ferromagnet FePd$_2$Te$_2$ (T$_C$ = 183K), which hosts 1D spin chains and strong in-plane anisotropy. Our first-principles calculations reveal highly anisotropic magnetic exchange interactions, confirming its 1D ferromagnetic nature. We modulate this behavior by Co and Ni substitution and introduce two new members of this family, CoPd$_2$Te$_2$ –a ferromagnet– and NiPd$_2$Te$_2$. Our results unveil the microscopic mechanisms governing the behaviour of FePd$_2$Te$_2$ and CoPd$_2$Te$_2$. Furthermore, we also demonstrate that the variation of the chain length is key to modulate magnetism. Finally, we determine the magnon dispersion, showcasing a pronounced anisotropy that enables unidirectional magnon propagation.

KEYWORDS: 1D magnetism, 2D materials, first-principles calculations, doping, straintronics.

The discovery of intrinsic magnetism in two-dimensional (2D) systems has revolutionized the fields of condensed matter, thin-films chemistry and material science, positioning 2D van der Waals (vdW) magnets as promising candidates for next-generation quantum technologies.[1–3] To date, a wide variety of 2D magnetic materials with a broad range of electronic and magnetic properties has been reported, including antiferromagnetic (AF) semiconductors such as CrSBr, $CrI_3$ and the $MPS_3$ family,[4–7] or the metallic ferromagnets $Fe_3GaTe_2$ and $Fe_xGeTe_2$ (x = 3, 4, 5).[8–12] Most of these systems exhibit symmetric arrangements of their magnetic centres, leading to spin-spin interactions in different directions and thus limiting the unidirectional control over their magnetic properties.[13,14] In this context, one-dimensional (1D) magnetism offers a compelling route to localize magnetism by minimizing interactions between neighbouring spins, allowing the emergence of exotic phenomena.[15–18] However, the practical isolation of materials hosting individual 1D spin chains possesses a major challenge, mainly due to their poor stability and fabrication complexity.[19,20] Thus, the rational design of novel 2D magnets that combine the strong anisotropy of 1D spin chains with the ultrathin 2D nature of these systems is of great scientific interest.

A prominent example of this class of systems is $AgCrP_2S_6$,[21,22] an anisotropic vdW antiferromagnet ($T_N$ = 21K) that shares structural similarities with the $MPS_3$ family.[14,23] Unlike the honeycomb magnetic lattice in $MPS_3$, $AgCrP_2S_6$ contains Cr atoms forming well-isolated 1D zigzag chains separated by non-magnetic Ag atoms, thus leading to a 1D magnetic behaviour within each layer. More recently, $FePd_2Te_2$ has positioned as a promising 2D vdW magnet hosting 1D spin chains, owing to its ferromagnetic (FM) order with a relatively high $T_C$ of 183K.[24] This 2D material crystallizes in a layered structure with Fe atoms forming zigzag chains along the crystallographic *b*-axis, displaying metallic conductivity. Additionally, it shows strong in-plane uniaxial magnetic anisotropy and a large coercive field, highlighting its potential for spintronics and magnonics applications.

One of the key advantages of low-dimensional magnetic materials is their tunability, as external perturbations such as strain or chemical doping can strongly influence their magnetic behaviour.[25–28] For instance, Co doping in $Fe_3GeTe_2$ modulates its magnetic anisotropy and pins domain walls,[29] while in $Fe_4GeTe_2$ it induces a transition from itinerant ferromagnetism to antiferromagnetism, accompanied by a reduction of its critical temperature.[30] Similarly, in $Fe_5GeTe_2$, Co substitution raises its $T_C$, reorients easy axis of magnetization, and triggers a transition from FM to AF order.[31,32] In $Fe_3GaTe_2$, Co doping also drives an AF-FM transition, ultimately leading to a spin glass state, the latter also observed upon addition of Ni atoms.[33,34] Therefore, given the potential of $FePd_2Te_2$ as a host for 1D spin chains, combined with the proven effectiveness of Co and Ni doping in tuning the magnetic properties of Fe-based metallic ferromagnets, a detailed investigation of the magnetic properties of $FePd_2Te_2$ along with the impact Co and Ni substitution deserves urgent attention.

In this work, we employ first-principles calculations to investigate the electronic and magnetic properties of $FePd_2Te_2$, unveiling the 1D nature of its exchange interactions and the mechanisms stabilizing its long-range ferromagnetism. We demonstrate that doping with Co and Ni atoms effectively tunes its magnetic properties, predict the dynamical stability of the isostructural $CoPd_2Te_2$ and $NiPd_2Te_2$ compounds and explore the potential of strain engineering to modulate their magnetic behaviour. Finally, we determine the magnon dispersion of $FePd_2Te_2$ and $CoPd_2Te_2$, showing that the pronounced anisotropy in their magnon spectra supports unidirectional propagation of magnons along their 1D spin chains.

FePd$_2$Te$_2$ is a recently synthesized quasi-2D crystal, which possesses 1D Fe zigzag chains along the *b* crystallographic axis and a relatively high T$_C$ of 183K.[24] It displays a monoclinic symmetry within the P2$_1$/*m* (N$_0$. 11) space group and lattice parameters of *a* = 7.50 Å, *b* = 3.95 Å, *c* = 7.74 Å, *α* = *γ* = 90°, and *β* = 118.15° (see Figure 1a). Firstly, we perform density-functional theory (DFT) calculations on FePd$_2$Te$_2$, obtaining optimized lattice parameters of 7.56 Å, *b* = 3.98 Å, *c* = 7.85 Å, and *α* = *γ* = 90°, and *β* = 118.08°, with a distance between adjacent 1D spin chains of 7.96Å, in good agreement with experimental results. The computed magnetic moment for each Fe atom is 2.75 μ$_B$, closely matching the reported value of 2.83 μ$_B$ for a single domain crystal,[24] while Pd and Te atoms exhibit null spin polarization. From the calculation of the electronic band structure, one can observe that the system is metallic (Figure 1c), with d orbitals of Fe playing a crucial role around the Fermi level and a significant contribution of d and p orbitals of Pd and Te atoms, respectively (Figure S1). The dynamic stability of bulk FePd$_2$Te$_2$ is assessed by performing phonon calculations. As observed in Figure 1d, the phonon dispersion presents no imaginary frequencies, indicating that the structure is stable.

Additionally, we calculated its formation energy (E$_{form}$) using the following expression[35]:

$$E_{form} = \frac{E_{FePd2te2} - 2xE_{Fe} - 4xE_{Pd} - 4xE_{Te}}{10}$$

where E$_{FePd2Te2}$ is the energy of FePd$_2$Te$_2$ and E$_{Fe}$, E$_{Pd}$ and E$_{Te}$ represent the energies of isolated Fe, Pd and Te atoms. We obtain E$_{form}$ = -0.53 eV/atom, indicating that the system is stable. Furthermore, we investigated the structural, electronic and magnetic properties of bulk isostructural CoPd$_2$Te$_2$ and NiPd$_2$Te$_2$. Both materials retain the monoclinic symmetry. The optimized lattice parameters of CoPd$_2$Te$_2$ are *a* = 7.53 Å, *b* = 4.00 Å, *c* = 7.66 Å, *α* = *γ* = 90°, and *β* = 115.53°, whereas for NiPd$_2$Te$_2$ they are *a* = 7.45 Å, *b* = 3.97 Å, *c* = 7.73 Å, *α* = *γ* =

90°, and $\beta$ = 115.17°. Our results reveal that CoPd$_2$Te$_2$ does possess a FM ground state, with each Co atom carrying a magnetic moment of 1.25 $\mu_B$. Conversely, NiPd$_2$Te$_2$ is paramagnetic. This is equivalent to the behaviour of Ni$_x$GeTe$_2$ (x = 3 and 5), which exhibits temperature-independent paramagnetism, in contrast to the itinerant ferromagnetism observed in Fe$_x$GeTe$_2$.[36,37] Regarding the electronic properties, the band structure evidences that both CoPd$_2$Te$_2$ and NiPd$_2$Te$_2$ display a metallic behaviour. Similar to FePd$_2$Te$_2$, the d orbitals of Co and Ni atoms contribute significantly near the Fermi level (Figures S2-3). Then, we assess their dynamical stability by means of their phonon spectra, which reveal no imaginary frequencies for CoPd$_2$Te$_2$ and a small negative contribution around Γ for NiPd$_2$Te$_2$ (Figure 1d). Moreover, they exhibit E$_{form}$ values of –0.42 eV/atom and –0.57 eV/atom, respectively, which are comparable to that of FePd$_2$Te$_2$, indicating the feasibility of synthesizing those compounds. Subsequently, we calculate the magnetic anisotropy energy (MAE) of FePd$_2$Te$_2$ and CoPd$_2$Te$_2$ from the total energy difference for spin orientations along the [101], [010], and [10$\bar{1}$] directions. In agreement with experimental results, FePd$_2$Te$_2$ presents uniaxial anisotropy, with the easy axis of magnetization oriented along the in-plane [101] direction. The [10$\bar{1}$] and [010] directions correspond to intermediate and hard axes, exhibiting MAE values of 0.52 meV/Fe and 1.08 meV/Fe, respectively. For CoPd$_2$Te$_2$, the magnetization aligns along the [101] direction, and the intermediate and hard axes are reversed relative to FePd$_2$Te$_2$, with corresponding MAE values of 0.25 meV/Co along [010] and 0.44 meV/Co along [10$\bar{1}$].

Next, we derive the magnetic exchange interactions for FePd$_2$Te$_2$ and CoPd$_2$Te$_2$ using a spin Hamiltonian with the following form:

$$H = -\sum_{i \neq j} J_{ij} \vec{S}_i \cdot \vec{S}_j - \sum_i A(\vec{S}_i^z)^2$$

where $J_{ij}$ denotes the isotropic exchange interaction between magnetic moments $S_i$ and $S_j$, and A represents the single-ion magnetic anisotropy constant.

The magnetic interaction picture in $FePd_2Te_2$ and $CoPd_2Te_2$ can be described by two primary in-plane contributions: (i) inter-site interactions between $Fe_1$ and $Fe_2$ sites, denoted as $J_{12}$, and (ii) interactions between equivalent magnetic centres, $Fe_1$–$Fe_1$, labelled as $J_{11}$. In constructing the spin Hamiltonian, the selected couplings are considered as a function of distance up to a maximum of 14 Å, beyond which they vanish. Analysing $J_{12}$, we observe that both $FePd_2Te_2$ and $CoPd_2Te_2$ exhibit FM behaviour (Figure 2c). Specifically, $FePd_2Te_2$ displays a value of 32.8 meV, which is reduced to 11.6 meV in its Co counterpart. In $FePd_2Te_2$, the strong FM character of $J_{12}$ is primarily governed by the $d_{xy}$-$d_{xy}$, $d_{xy}$-$d_{z2}$, $d_{z2}$-$d_{z2}$, $d_{x2-y2}$-$d_{x2-y2}$ and $d_{xz}$-$d_{yz}$ orbitals (Figure 2d). This FM behaviour is partially diminished by AF contributions from $d_{x2-y2}$-$d_{xz}$ and $d_{x2-y2}$-$d_{yz}$ orbitals. For $CoPd_2Te_2$, the FM nature of $J_{12}$ is attributed to the contribution of $d_{xy}$-$d_{xy}$ and $d_{xy}$-$d_{z2}$ orbitals. Figure 2d illustrates the difference in the orbital-resolved contribution to $J_{12}$ ($\Delta J_{12}$) between both materials, obtained by subtracting the contribution of $CoPd_2Te_2$ from $FePd_2Te_2$. This reveals that the enhanced ferromagnetism in $FePd_2Te_2$ arises from stronger FM character of $d_{x2-y2}$-$d_{x2-y2}$, $d_{z2}$-$d_{z2}$ and $d_{xz}$-$d_{yz}$ relative to $CoPd_2Te_2$. A distinct trend is observed for $J_{11}$ between both systems (Figure 2c). In $CoPd_2Te_2$, $J_{11}$ is FM, whereas in $FePd_2Te_2$ it is AF, primarily due to contributions of $d_{z2}$-$d_{z2}$ orbitals (Figure 2e). The coexistence of competing FM $J_{12}$ and AF $J_{11}$ interactions in $FePd_2Te_2$ leads to a geometrically frustrated spin system, thereby partially suppressing the net ferromagnetism within the lattice, similar to the behaviour of $Fe_3GeTe_2$.[38,39] Nevertheless, the overall magnetic ordering within $FePd_2Te_2$ is FM, as the dominant magnitude of $J_{12}$ (32.8 meV) outweighs the weaker AF contribution from $J_{11}$ (-3.3 meV). In the studied range, we also include $J_{12}$ and $J_{11}$ couplings that encompass interchain interactions. The interactions between nearest interchain neighbours have values of $J_{12}$ = 0.1 and 0.06 meV as well as $J_{11}$ = 0.07 and 0.15 meV for $FePd_2Te_2$ and $CoPd_2Te_2$, respectively, demonstrating the intrachain 1D nature of magnetic couplings. Interlayer couplings between adjacent layers were also computed, revealing an overall FM character in both materials. However, their magnitudes are considerably smaller compared to dominant intrachain interactions (Tables S1-4). The calculated $T_C$ for $FePd_2Te_2$ is 240K, which is not too far from the experimental value of 183K. On the other hand, $CoPd_2Te_2$ is predicted to exhibit a lower $T_C$ of 140K, consistent with its weaker $J_{12}$ exchange interaction.

We further investigated the magnetic behaviour of $FePd_2Te_2$ upon partial atomic substitution, where 50% of the Fe atoms along each 1D chain are replaced by either Co or Ni, resulting in the compounds $Fe_{0.5}Co_{0.5}Pd_2Te_2$ and $Fe_{0.5}Ni_{0.5}Pd_2Te_2$, respectively. For such purpose, two different atomic configurations were examined. In the first one, referred as *ALT*, Fe atoms are substituted by Co (or Ni) in an alternating manner (Figures 3a,b), with each Fe atom surrounded by two Co atoms. In the second arrangement, two consecutive Fe atoms are followed by two Co atoms, which we refer as *CON*. This results in each Fe atom being coordinated by one neighbouring Fe and one Co (Figures 3d,e). Both structures were optimized, yielding similar lattice parameters, where the $P2_1/m$ symmetry is retained (Table S5). A comparison between *ALT* and *CON* configurations reveals that the latter is energetically more favourable for both materials (Table S6). Regarding the magnetic properties, Table 1 summarizes the values of magnetic moments and MAE for these compounds. $Fe_{0.5}Co_{0.5}Pd_2Te_2$ in the *ALT* configuration, the magnetic moments are 2.80 $\mu_B$ for Fe and 1.53 $\mu_B$ for Co, the latter enhanced relative to pristine $CoPd_2Te_2$ (1.26 $\mu_B$) owing to the coordination with two neighbouring Fe atoms. In the *CON* configuration, magnetic moments of Fe and Co slightly decrease to 2.78 $\mu_B$ and 1.37 $\mu_B$, respectively, as each Co atom is now coordinated with one Fe and one Co, exhibiting values closer to $CoPd_2Te_2$. For $Fe_{0.5}Ni_{0.5}Pd_2Te_2$, Ni atoms exhibit minimal polarization, with magnetic moments of 0.23 $\mu_B$ (*ALT*) and 0.18 $\mu_B$ (*CON*), consistent with those reported for Ni-doped $Fe_5GeTe_2$.[40] Regarding MAE, we observe a shift in the easy magnetic axis for $Fe_{0.5}Co_{0.5}Pd_2Te_2$ in the *ALT* configuration from [101] to the [10$\bar{1}$] direction, with a corresponding MAE of -52.1 µeV/magnetic atom, therefore stabilizing an out of plane magnetic ground state. This resembles the behaviour exhibited by Co-doped $Fe_5GeTe_2$, where the addition of Co atoms modifies its magnetic easy axis.[31,32] Conversely, in *CON* configuration, the system retains the magnetization along [101], requiring 0.16 meV to reorient spins towards either [101] or [10$\bar{1}$].

Table 1. Values of $\mu_B$ and MAE for $FePd_2Te_2$, $CoPd_2Te_2$, $Fe_{0.5}Co_{0.5}Pd_2Te_2$ and $Fe_{0.5}Ni_{0.5}Pd_2Te_2$.

| Material | Arrangement | $\mu_B$ (Fe) | $\mu_B$ (Co) | $\mu_B$ (Ni) | [10$\bar{1}$] − [101] (meV/magnetic atom) | [010] − [101] (meV/magnetic atom) |
|---|---|---|---|---|---|---|
| $FePd_2Te_2$ | − | 2.75 | − | − | 0.52 | 1.08 |
| $Fe_{0.5}Co_{0.5}Pd_2Te_2$ | ALT | 2.80 | 1.53 | − | -0.052 | 0.047 |
|  | CON | 2.78 | 1.37 | − | 0.16 | 0.16 |
| $CoPd_2Te_2$ | − | − | 1.26 | − | 0.44 | 0.25 |
| $Fe_{0.5}Ni_{0.5}Pd_2Te_2$ | ALT | 2.80 | − | 0.23 | 0.77 | 0.64 |
|  | CON | 2.78 | − | 0.18 | 0.49 | 0.21 |

For $Fe_{0.5}Ni_{0.5}Pd_2Te_2$, the magnetic easy axis remains unchanged. However, the MAE is lower in the *CON* configuration (0.49 meV/magnetic atom) compared to *ALT* (0.77 meV/magnetic atom), likely due to the presence of two consecutive non-magnetic Ni atoms per chain, which reduces the overall anisotropy.

We computed the magnetic exchange couplings and performed atomistic simulations to estimate the $T_C$. In the *ALT* configuration of $Fe_{0.5}Co_{0.5}Pd_2Te_2$, there is a strong FM coupling between $Fe_1$ and $Co_2$ atoms (18.8 meV). This is notably reduced compared to the $Fe_1$-$Fe_2$ interaction in $FePd_2Te_2$ ($J_{12}$ = 32.8 meV), primarily due to the reduced magnetic moment of Co relative to Fe. Interactions between $Fe_1$-$Fe_1$ and $Co_2$-$Co_2$ retain their AF and FM characters, respectively, as in the pristine compounds. The estimated $T_C$ for $Fe_{0.5}Co_{0.5}Pd_2Te_2$ is 150K. A similar behaviour is observed in $Fe_{0.5}Ni_{0.5}Pd_2Te_2$, where the FM coupling $Fe_1$-$Ni_2$ is 5 meV. Furthermore, AF interaction between $Fe_1$ atoms is significantly suppressed and the $Ni_2$-$Ni_2$ coupling is negligible due to the minimal polarization of Ni atoms. As a result, the estimated $T_C$ is 100K, primarily driven by the $Fe_1$-$Ni_2$ interaction.

According to our calculations, in the *CON* configuration for $Fe_{0.5}Co_{0.5}Pd_2Te_2$, there is a robust FM coupling of 27 meV between $Fe_1$ and $Fe_2$ sites along with more subtle FM couplings between $Co_1$-$Fe_2$ and $Co_1$-$Co_2$. Unlike *ALT*, the *CON* configuration lacks short-range AF couplings between $Fe_1$-$Fe_1$. In contrast, its equivalent $Fe_1$-$Co_1$ exchange is FM, showing a modest value of 1meV (Figure S5). Due to the enhanced FM character of the mentioned couplings and the suppression of competing AF interactions, the estimated $T_C$ increases from 150K in *ALT* to 170K in *CON*. For $Fe_{0.5}Ni_{0.5}Pd_2Te_2$, the $Fe_1$-$Fe_2$ FM coupling is 35 meV and the $Ni_1$-$Fe_2$ interaction is 5 meV. However, the $Ni_1$-$Ni_2$ coupling is effectively zero, consistent with the near-zero magnetic moment of Ni atoms. Although the overall ferromagnetism is enhanced compared to the *ALT* configuration, the system does not retain long-range magnetic ordering ($T_C$ = 0K) due to the absence of short-range Ni-Ni interactions. Therefore, we foresee that the experimental incorporation of consecutive Ni atoms into the $FePd_2Te_2$ lattice would suppress the net ferromagnetism along the 1D chains. Additionally, we computed the energy difference between interlayer AF and FM configurations for both compounds (Table S7). Our results indicate that the interlayer ferromagnetism is always retained, differently from the Co-induced antiferromagnetism observed in $Fe_4GeTe_2$, $Fe_5GeTe_2$ and $Fe_3GaTe_2$.[30–34]

Then, we examine the magnetic behaviour of $FePd_2Te_2$ and $CoPd_2Te_2$ under uniaxial strain along the intrachain [010] direction. The magnetic moments of Fe and Co atoms exhibit subtle variations under deformation (Figure S6). Regarding anisotropy, Figures 4a,d indicate that the easy axis of magnetization is unaltered across the entire range, where MAE remains largely unaffected in $FePd_2Te_2$. By contrast, it increases under tensile strain in $CoPd_2Te_2$. From the evolution of magnetic couplings, one can observe that $J_{12}$ becomes more FM under compression, reaching 35.8 meV (-2%) for $FePd_2Te_2$ and 11.7 meV (-1.5%) for $CoPd_2Te_2$ (Figures 4b,e). Conversely, $J_{11}$ becomes more AF in $FePd_2Te_2$ due to an enhanced contribution from the $d_{z^2}$-$d_{z^2}$ mechanism (Figure S7). Additionally, we computed the evolution of long-range exchange interactions, namely $J_{long}$. These account for the sum of all the couplings beyond the nearest neighbours $J_{12}$ and $J_{11}$. In both $FePd_2Te_2$ and $CoPd_2Te_2$, $J_{long}$ increases upon tensile strain. This, together with the observed trend of $J_{11}$ results in an enhancement of $T_C$ upon tensile strain and a decrease upon compression (Figures 4c,f).

The analysis of the magnon spectra for $FePd_2Te_2$ and $CoPd_2Te_2$ reveals significant anisotropy (Figure 5) consistent with the previously discussed exchange interactions. In both compounds, strongly dispersive branches are observed along the $\Gamma-Z$ direction, corresponding to magnons propagating along the 1D spin chains. In contrast, dispersion is notably weaker along directions perpendicular to them (e.g., $\Gamma-B$ or $\Gamma-Y_2$), indicating suppressed interchain magnon transport. Under +2% tensile deformation, $FePd_2Te_2$ exhibits an ~8% reduction in the energy of optical magnons at $\Gamma$, attributed to the strain-induced weakening of $J_{12}$. Conversely, the energy of acoustic magnons increase with strain, reflecting the enhancement of $J_{long}$. The same trend is observed for $CoPd_2Te_2$; however, the optical magnons are marginally affected at $\Gamma$ due to the weaker strain dependence of $J_{12}$ in this compound.

In conclusion, we investigate the magnetic properties of $FePd_2Te_2$ using first principles calculations. Our results demonstrate that its magnetic behavior is predominantly governed by intrachain exchange interactions $J_{12}$ and $J_{11}$, underscoring the 1D magnetic character of this 2D material. We also propose two new isostructural compounds, namely $CoPd_2Te_2$ and $NiPd_2Te_2$, and confirm their dynamical stability. $CoPd_2Te_2$ exhibits a FM ground state, while $NiPd_2Te_2$ shows paramagnetic behavior. Furthermore, we provide a detailed microscopic analysis of the magnetic exchange interactions in $FePd_2Te_2$ and $CoPd_2Te_2$, revealing the distinct mechanisms underlying their magnetic behavior and show the potential of Co and Ni substitution to modulate the exchange couplings and anisotropy of $FePd_2Te_2$. Finally, we demonstrate that mechanical strain offers an effective route for tuning the magnetic properties of $FePd_2Te_2$ and $CoPd_2Te_2$ and show the pronounced anisotropy in their magnon spectra supports directional propagation of magnons along their 1D spin chains. Our findings highlight the potential of two-dimensional materials hosting 1D chains, and how their chemical and structural manipulation offers a promising platform for the design of magnetic and spintronic devices with tailored anisotropic characteristics.

METHODS

DFT calculations on $FePd_2Te_2$, $CoPd_2Te_2$ and $NiPd_2Te_2$ were performed using the VASP package[41] using the generalized gradient approximation (GGA). We relaxed both the atomic coordinates and lattice parameters employing a plane-wave cutoff of 400 eV. The Brillouin zone was sampled using a 8×10×8 k-point Monkhorst–Pack that was increased to 10×12×10 for the calculations of MAE. The phonon spectra for the three compounds were obtained using the Phonopy package[42] with supercells of dimensions 2×2×2. The magnetic couplings were evaluated as a function of distance until saturation using the TB2J software[43] and employing supercells of dimensions 20×40×20. A tight-binding Hamiltonian served as input for these calculations, obtained via the Wannier90 code,[44] with a reduced basis set formed by *d* orbitals of Fe, Co and Ni as well as the *d* orbitals of Pd and *p* orbitals of Te. Simulations of $T_C$ were performed by means of the VAMPIRE code,[45] with 10000 steps for equilibration and averaging phases along with supercells of dimensions 15 nm ×15 nm ×15 nm.

AUTHOR INFORMATION

**Corresponding Author**

*E-mail: j.jaime.baldovi@uv.es.

**Author Contributions**

This work is part of the PhD thesis of A.M.R. All authors have given approval to the final version of the manuscript.

**Notes**

The authors declare no competing financial interest.

ACKNOWLEDGMENTS

The authors acknowledge the financial support from the European Union (ERC-2021-StG101042680 2D-SMARTiES) and the Excellence Unit "María de Maeztu" CEX2019-000919-M). J.J.B acknowledges the Generalitat Valenciana (grant CIDEXG/2023/1) and A.M.R. thanks the Spanish MIU (Grant No FPU21/04195). The calculations were performed on the Tirant III cluster of the Servei d'Informática of the University of València.

FIGURES

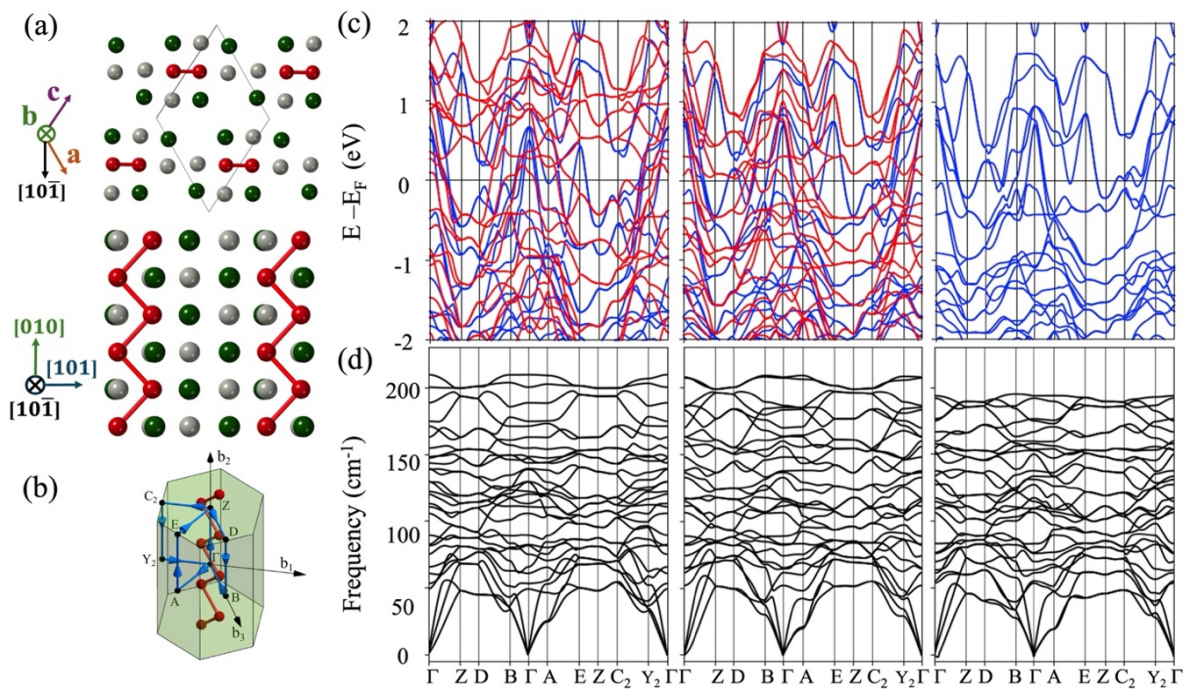

Figure 1. (a) Top view of bulk $FePd_2Te_2$. Color code: Fe (red), Pd (grey) and Te (green). (b) 3D Brillouin zone (BZ) of $FePd_2Te_2$ and illustration of 1D chain of the material. (c) Band structure and (d) phonon spectra for $FePd_2Te_2$, $CoPd_2Te_2$ and $NiPd_2Te_2$ (from left to right, respectively).

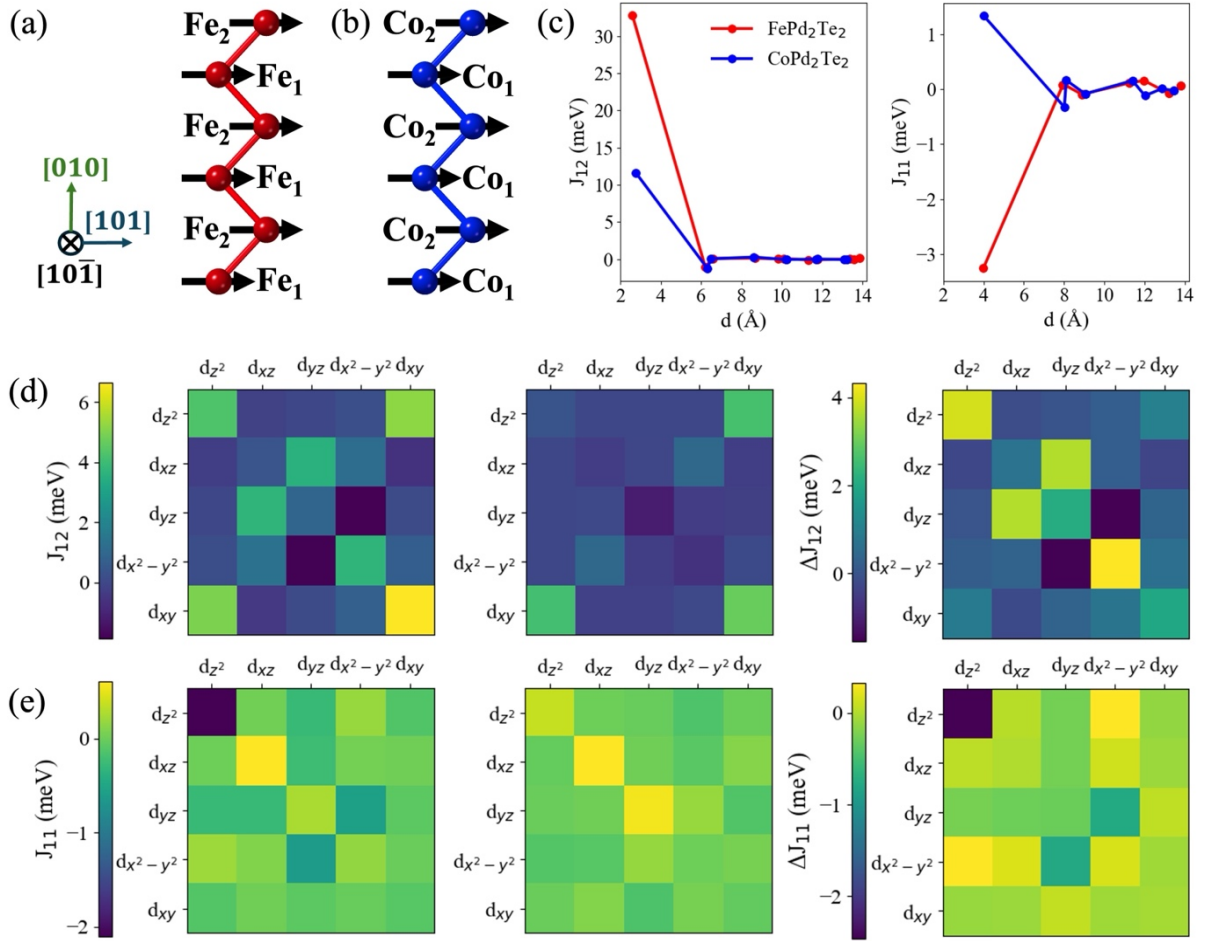

Figure 2. Top view of 1D chains of (a) $FePd_2Te_2$ and (b) $CoPd_2Te_2$. For simplicity, we only depict the magnetic centres Fe and Co atoms. (c) Exchange interactions $J_{12}$ and $J_{11}$ for $FePd_2Te_2$ (red) and $CoPd_2Te_2$ (blue). (d) Orbital-resolved $J_{12}$ exchange coupling for $FePd_2Te_2$ (left), $CoPd_2Te_2$ (middle) and subtracted contribution, $\Delta J_{12}$ (right). (e) Same as (d), but for $J_{11}$.

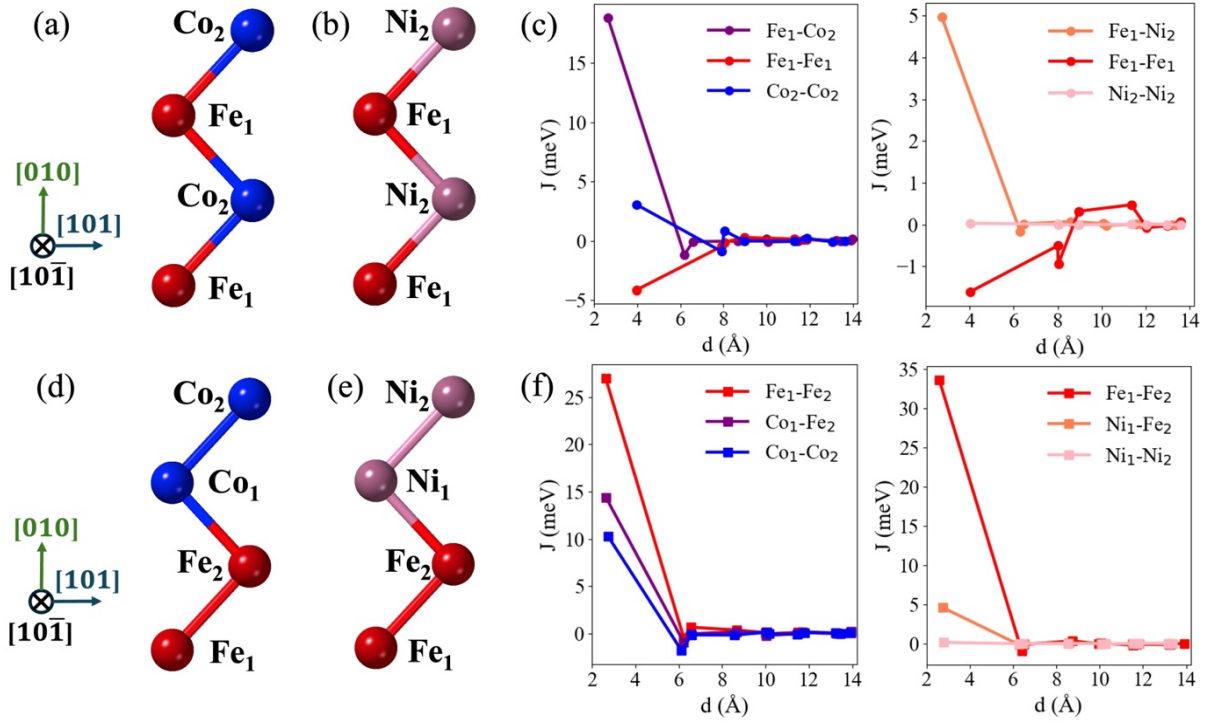

Figure 3. Top view of 1D chains in the *ALT* configuration for (a) $Fe_{0.5}Co_{0.5}Pd_2Te_2$ and (a) $Fe_{0.5}Ni_{0.5}Pd_2Te_2$ along with their corresponding (c) magnetic couplings for each material. Top view of 1D chains in the *CON* configuration for (d) $Fe_{0.5}Co_{0.5}Pd_2Te_2$ and (e) $Fe_{0.5}Ni_{0.5}Pd_2Te_2$ along with their corresponding (f) magnetic couplings for each material.

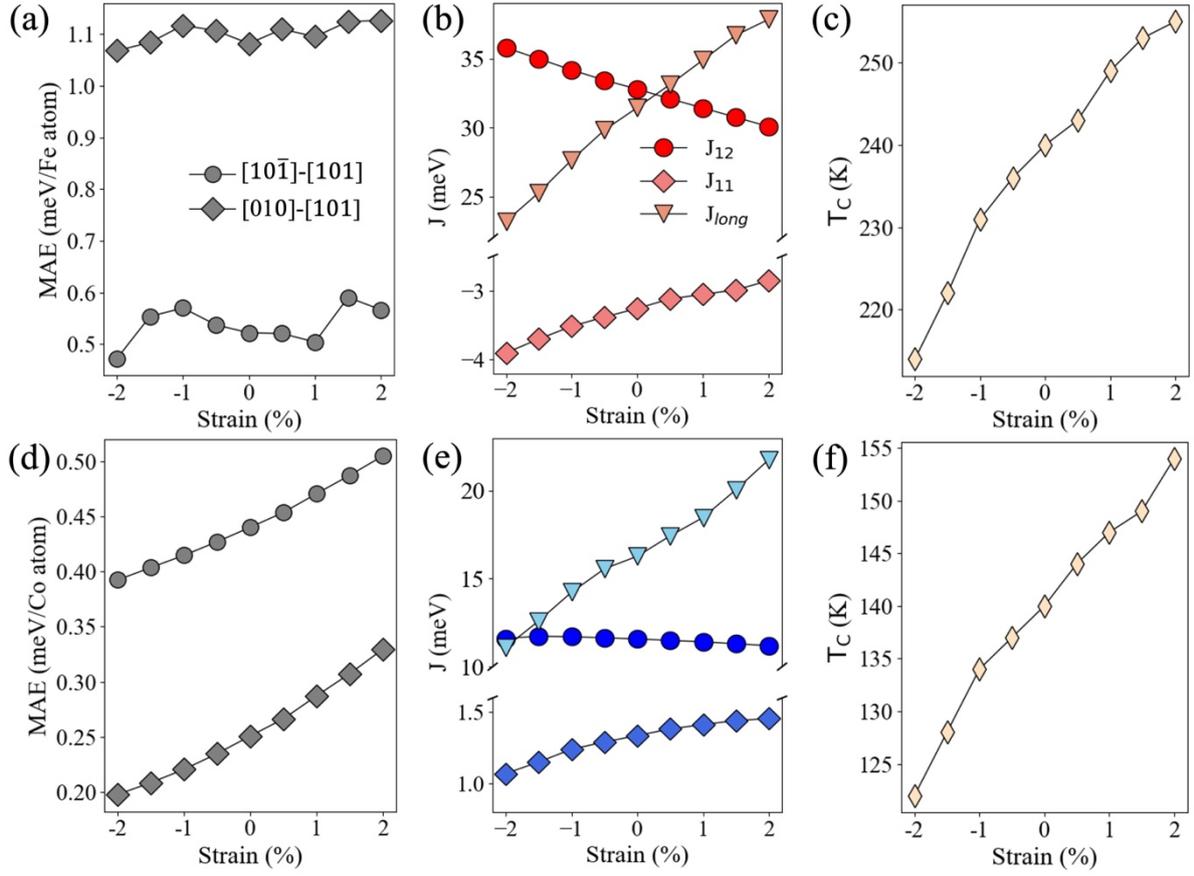

Figure 4. Evolution of (a) MAE, (b) exchange couplings $J_{12}$, $J_{11}$ and $J_{12}$ and (c) $T_C$ for $FePd_2Te_2$ upon applied strain along the [010] direction; and corresponding evolution of (d) MAE, (e) $J_{12}$, $J_{11}$ and $J_{12}$ and (f) $T_C$ for $CoPd_2Te_2$.

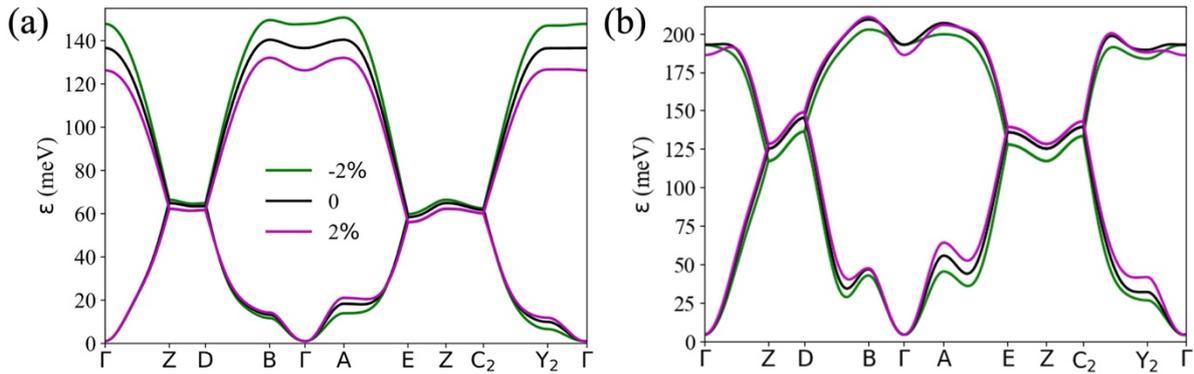

Figure 5. Magnon dispersion for (a) $FePd_2Te_2$ and (b) $CoPd_2Te_2$ as a function of strain applied along the [010] direction.